\documentclass[conference]{IEEEtran}
\IEEEoverridecommandlockouts
\usepackage{cite}
\usepackage{url}
\usepackage{amssymb,amsfonts}
\usepackage{amsmath}
\usepackage{algorithm}
\usepackage{algpseudocode}
\usepackage{graphicx}
\usepackage{textcomp}
\usepackage{xcolor}
\usepackage{fancyhdr}
\usepackage{subfigure}
\usepackage[left=1.62cm,right=1.62cm,top=1.9cm, bottom=2.67cm]{geometry}
\def\BibTeX{{\rm B\kern-.05em{\sc i\kern-.025em b}\kern-.08em
    T\kern-.1667em\lower.7ex\hbox{E}\kern-.125emX}}

\newboolean{showcomments}
\setboolean{showcomments}{true}
\ifthenelse{\boolean{showcomments}}
{
}

\newcommand*{\affmark}[1][*]{\textsuperscript{#1}}

\begin{document}

\title{LLM-Based Intent Processing and Network Optimization Using Attention-Based Hierarchical Reinforcement Learning}

\author{\IEEEauthorblockN{Md~Arafat~Habib\affmark[1], Pedro Enrique Iturria Rivera\affmark[1], Yigit Ozcan\affmark[2], Medhat Elsayed\affmark[2], \\Majid Bavand\affmark[2], Raimundus Gaigalas\affmark[3] and  Melike Erol-Kantarci\affmark[1], \IEEEmembership{Senior Member,~IEEE}}
\IEEEauthorblockA{\affmark[1]\textit{School of Electrical Engineering and Computer Science, University of Ottawa, Ottawa, Canada}}  \affmark[2]\textit{Ericsson Inc., Ottawa, Canada} \affmark[3]\textit{Ericsson AB, Stockholm, Sweden}\\
Emails:\{mhabi050, pitur008, melike.erolkantarci\}@uottawa.ca, \\\{yigit.ozcan, medhat.elsayed, majid.bavand, raimundas.gaigalas \}@ericsson.com \vspace{-1em}}

\maketitle

\thispagestyle{fancy}   
\fancyhead{}                
\lhead{Accepted by the WCNC 2025. The final version of the published material can change at any time without any notice. Personal use is permitted only. \copyright2025 IEEE}
\cfoot{}
\renewcommand{\headrulewidth}{0pt} 

\begin{abstract}
Intent-based network automation is a promising tool that enables easier network management; however, certain challenges must be addressed effectively. These are: 1) processing intents, i.e., identification of logic and necessary parameters to fulfill an intent, 2) validating an intent to align it with current network status, and 3) satisfying intents via network optimizing applications. This paper addresses these points via a three-fold strategy to introduce intent-based automation for modern 5G architectures. First, intents are processed via a lightweight Large Language Model (LLM). Secondly, once an intent is processed, it is validated against future incoming traffic volume profiles (high or low). Finally, a series of network optimization applications has been developed. With their machine learning-based functionalities, they can improve certain key performance indicators such as throughput, delay, and energy efficiency. In the final stage, using an attention-based hierarchical reinforcement learning algorithm, these applications are optimally initiated to satisfy the intent of an operator. Our simulations show that the proposed method can achieve at least a $12\%$ increase in throughput, a $17.1 \%$ increase in energy efficiency, and a $26.5\%$ decrease in network delay compared to the baseline algorithms. 

\end{abstract}


\section{Introduction}
\label{s1} 

An intent can be defined as the expectations, including requirements, goals, and constraints for a specific service or network management workflow\cite{20}. It is typically understandable by humans and must be interpreted by machines without ambiguity. Furthermore, it emphasizes ``What'' needs to be achieved, rather than focusing on ``How'' to achieve the outcomes, specifying the desired metrics without detailing the process to attain them \cite{20}. It is anticipated that Mobile Network Operators (MNOs) will leverage advancements in natural language processing to express intents in plain language for automated network management. In modern 5G networks, achieving such intent-based automation can be challenging especially because intents can vary widely based on network conditions, customer demands, and industry-specific terminologies \cite{20}. Therefore, a contextual understanding of the intents is vital for accurate intent processing. Large Language Models (LLMs) with its transformer architecture excel at capturing such context and understanding the underlying meaning of sentences (intents) within a broader spectrum \cite{9}. 

Processing intents expressed in natural language by the human network operator is important to derive policies for network performance optimization. However, allowing any kind of intent to put an impact on the network configurations can lead to serious degradation in performance. For example, an intent: ``Increase energy efficiency by $30\%$" during peak hours can degrade performance due to high incoming traffic volume. This kind of scenario requires a sophisticated method of intent validation using contemporary network status to ensure a seamless user experience without compromising any QoS.

The concept of an Open Radio Access Network (Open RAN) facilitates the idea of disaggregating the traditional RAN and promoting interoperability between components from different vendors. This concept enables the deployment of network-optimizing applications (apps) in intelligent RAN controllers, which can be triggered by processed intents. In \cite {1}, intents are transformed into goals for a Hierarchical Reinforcement Learning (HRL) algorithm to initiate such applications for intent fulfillment. However, the computational intensity of selecting optimal combinations from numerous applications becomes significant for an AI agent.   

To this end, we propose a three-fold strategy to perform intent processing, validation, and application initiation based on intents. We begin by processing intents using a pre-trained LLM named Bidirectional Encoder Representations from Transformers (BERT) \cite{5}. In the next step, intent validation is performed using a Transformer-based time series predictor for upcoming traffic volume \cite{8}. This ensures that the intended adjustments to the network are not only theoretically sound but also practical and feasible given expected future traffic patterns. Lastly, to overcome the computational burden of selecting the appropriate combination of applications from a high number of possible choices, we utilize an attention-based HRL framework that takes the processed intents (magnitude of the intended performance metric by an operator) as goals. By using an attention mechanism to filter only the feasible options \cite{7}, the system can significantly reduce the computational burden.

The simulation results demonstrate that the proposed scheme yields significant improvements compared to two baseline approaches: HRL without attention and intent validation, and a single-application scenario developed using Deep RL (DRL). The proposed approach surpasses the HRL baseline in terms of throughput, delay, and energy efficiency by $12.02\%$, $26.5\%$, and $17.1\%$ respectively. Moreover, compared to the DRL baseline, the proposed method achieves superior results across the same performance metrics, with improvements of $17.2\%$, $48.6\%$, and $39.3\%$ respectively.        

\section{Related work}
\label{s2}

Even though using LLMs to process intents for network optimization is a relatively new concept, a few works have been found in the literature. Wang et al.'s development of a transformer model based on an LLM is one such example, where multi-modal representation learning is employed to integrate various forms of network data into a unified feature space \cite{4}. Similarly, Kristina et al. have introduced a pipeline that utilizes an LLM to process intents into structured, policy-based abstractions, linking them to APIs for execution \cite{9}. 

To manage network applications based on operator intents, an approach based on HRL is proposed in \cite{1} where intents are processed into goals for the control algorithm to optimize performance. Polese et al. propose a framework to collect control requests (intents) and select the optimal Machine Learning (ML) models to achieve the operator’s goals to avoid conflicts \cite{2}.

Compared with existing literature, the main contribution of this work lies in deploying an LLM to interpret a human network operator's intent and extract specific parameters for further network performance optimization. Processed intents are validated against future wireless traffic conditions. Intents lead HRL agents to initiate and orchestrate multiple network applications with an efficient exploration of action space via an attention mechanism.  

\section{System Model}
\label{s3}

A cellular system employing downlink Orthogonal Frequency Division Multiplexing (OFDM) is considered in this work. There are $B$ base stations that concurrently serve $U$ users. Within the macro cell's coverage area, multiple small cells are deployed. The system supports $K$ classes of traffic, and users are connected through dual connectivity across multiple Radio Access Technologies (RATs). These RAT classes can facilitate various technologies like LTE, and 5G.

\begin{figure}[!t]
\centerline{\includegraphics[width=1\linewidth]{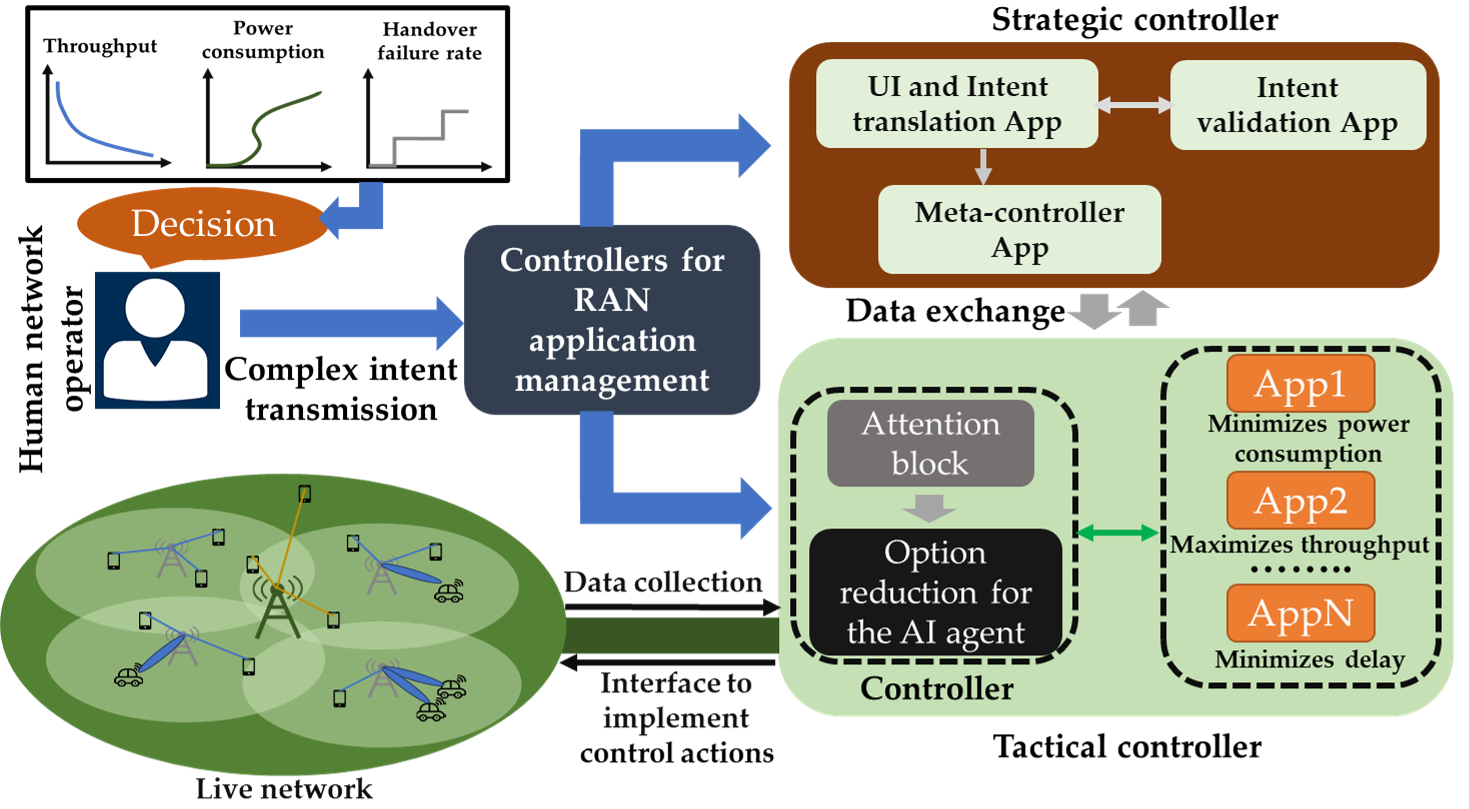}}
\caption{Three-step methodology for intent processing, validation, and performance optimization.}
\label{fig2}
\vspace{-1.2em}
\end{figure}

The system model, shown in Fig. \ref{fig2}, proposes two distinct controllers responsible for managing RAN functionalities: the strategic controller and the tactical controller. The strategic controller on top is responsible for long-term goals, such as network-wide performance optimizations, and policy formulation. It can define the network’s broader objectives, including service quality guarantees, energy efficiency, or overall traffic management for a timespan. The tactical controller on the bottom focuses on real-time decisions, such as handling immediate network changes, performing handovers, or dynamically adjusting resources based on the current network state. It operates within the framework set by the strategic controller but is more focused on near-real-time adjustments to ensure smooth network operation from moment to moment.

The controllers host applications focusing on control and optimization tasks across two different time scales. We deploy BSs capable of switching bands from 5G mid band to high band frequencies. This deployment enables us to cater to high-throughput traffic by employing intelligent beamforming techniques. An application can be developed to regulate power based on UE location, utilizing minimal transmission power for enhanced energy efficiency. The system facilitates analog beamforming, with each BS employing a uniform linear array of $\mu$ antennas \cite{12}. Beamforming weights for each beamforming vector are implemented using constant modulus phase shifters. Additionally, we assume the presence of a beam steering-based codebook $F$, from which beamforming vectors are selected \cite{12}. Each BS $b$ operates with a transmit power $P_{TX,b} \in P$, where $P$ represents the set of candidate transmit powers. The energy consumption model for the BS is obtained\ from \cite{13}.

\subsection{Problem Definition}

Based on the system and network model discussed so far, we introduce multiple applications. There are two sets of them. The first set consists of intent translation, validation, and meta-controller (associated with our RL-based control algorithm). The second set consists of traffic steering, beamforming, cell sleeping, power allocation, and handover management applications that can be directly controlled by the first set. All applications from the first set, along with cell sleeping, operate on a time frame longer than 1 second within the strategic controller, while the others preferably operate on a time scale between 10 ms and 1s within the tactical controller.

Let us define $\Psi$ as the collection of applications, and their various combinations. Let $\sigma$ represent the subset of $\Psi$ which contains at least one element that is capable of enhancing network performance in response to operator intents. We also define $\rho$ as the set of potential KPIs that an application might improve, and $Q_s$ as the set of QoS requirements that must be met by the system. Given these definitions, the challenge of orchestrating applications based on operator intent can be formulated as follows: 
\begin{equation} \label{eq4}
\begin{split}
     P_1: \max\sum_{\rho\in \varrho}\sum_{q_s\in Q_s}(C_\rho-\varpi \gamma_{q_s}), \quad\quad\quad  \\
     \text{s.t.} \quad \forall(\Psi)\exists(\sigma): \zeta(O)=1, \quad\quad \quad
\end{split}
\end{equation}
where $C$ is the magnitude of the intended performance metric the operator intends to improve. $\varpi$ is the penalty parameter for QoS requirement violation, and $\gamma_{q_s}$ represents the number of UEs whose QoS requirements have been violated. The evaluation is binary: a UE is counted only if its QoS standards are not met. Lastly, $\zeta(O)$ is the proposition that ``An application can improve a performance metric", which is either ‘0’ or ‘1’.        

The second problem we want to address in this work is associated with intent validation. QoS parameters deviate from the originally defined QoS metrics over time due to performance degradation in the wireless networks, causing QoS drifts. For a specific traffic class $k$ ($K = \{k_1, k_2, \ldots ,k_\eta\}$), with its set of QoS requirements $Q_s$ ($Q_s=\{Q_{s_1},Q_{s_2}, \ldots ,Q_{s_j}\}$) and corresponding performance metrics $C$, the QoS drift can be formulated as follows: 
\begin{equation} \label{eq5}
\begin{split}
     P_2: \min \sum_{q_s\in Q_s}\sum_{c\in C} \{D_{QoS} - A_{QoS}\}, \quad \quad \\
     \text{s.t.} \quad A_{QoS} \leq D_{QoS} \quad \text{and} \quad A_{QoS}>0.
\end{split}
\end{equation}

In (\ref{eq5}), $A_{QoS}$ represents the achieved value of a metric (e.g. delay, throughput) for a certain traffic type at time $t$, and $D_{QoS}$ represents the pre-defined value of the same parameter which has to be maintained for optimal performance. The objective here is to validate intents that will not result in QoS drifts. For cases where $A_{QoS} > D_{QoS}$, we consider there is no QoS drift and the intent is valid. Otherwise, we calculate the QoS drift, $I_{Q}=D_{QoS} - A_{QoS}$ based on (\ref{eq5}).

Another objective of this work is to reduce the large action space, which consists of numerous potential combinations of applications. To solve this problem, we define an attention function $\alpha(s_t, i, a)$ that scores each action $a \in A$ based on the current state $s_t$ and the operator's intent $i$. 

\section{Proposed Methodology}
\label{s4}
In this section, first, we discuss processing intents using an LLM and then predictive intent validation. The last part of this section consists of an HRL-based attention mechanism for application initiation and orchestration.

\subsection{Intent Processing using LLM}
Intent-based networking simplifies network and administrative tasks by automating system operations through the comprehension and execution of operator-defined intents, such as “Increase overall energy efficiency by 10\%”, “Boost system throughput by 15\%”, or “Reduce network delay by 13\%”. These intents focus on improving metrics like throughput, energy efficiency, and network delay. Our approach uses few-shot learning to classify intents and extract relevant keywords. The methodology involves constructing a prompt that includes a few examples of intents with their corresponding types and keywords. This prompt is then used to query the LLM, which classifies new intents and extracts the required information based on the provided examples. Before processing new intents, a set of examples is prepared. Each example consists of:
\begin{itemize}
    \item \textbf{Intent:} A natural language statement describing the intent.
    \item \textbf{Type:} The category or classification of the intent.
    \item \textbf{Keywords:} Key phrases or terms extracted from the intent.
\end{itemize} 

A prompt is constructed by sequentially adding the prepared examples. Each example is formatted to include the intent ($I=\{i_1,i_2, \ldots, i_n\}$), type ($I_{t_y}= \{t_{y_1}, t_{y_2},...,t_{y_m}\}$), and keywords ($\Lambda=\{\lambda_1, \lambda_2,\ldots,\lambda_w\}$). After adding all the examples, the new intent is appended to the prompt ($p$), followed by a request for classification and keyword extraction. The constructed prompt is sent to the pre-trained LLM (a lightweight version of BERT) \cite{14}. The LLM uses the context provided by the examples to understand the task and generate a response that includes the classification and keywords for the new intent. The response from the LLM is parsed to extract the classified type of the new intent and the extracted keywords. Each new intent's classified type and keywords are compiled into a result set.

\begin{algorithm}
\caption{Intent Processing Using LLM}
\begin{algorithmic}[1]
\Require $E = \{(i_1, t_{y_1}, \lambda_1), (i_2, t_{y_2}, \lambda_2),...\}$, $\Xi = \{(i, t_y, \lambda)\}$ 

\Procedure{ClassifyAndExtract}{$I, E, LLM$}
    \State $\Xi \gets \emptyset$ 
    \ForAll{$i \in I$}
        \State $p \gets $ \textsc{CreatePrompt}($i, E$)
        \State $response \gets $ \textsc{QueryLLM}($p, LLM$)
        \State $(t_y, \lambda) \gets $ \textsc{ParseResponse}($response$) 
        \State $\Xi \gets \Xi \cup \{(i, t_y, \lambda)\}$ 
    \EndFor
    \State \Return $\Xi$
\EndProcedure

\Procedure{CreatePrompt}{$i, E$}
    \State $p \gets ``\quad"$
    \ForAll{$e \in E$}
        \State $p \gets p + \text{``\textbf{Example:}"}$
        \State $p \gets p + \text{``\textbf{Intent:} "} + e[0]$
        \State $p \gets p + \text{``\textbf{Type:} "} + e[1]$
        \State $p \gets p + \text{``\textbf{Keywords:} "} + e[2]$
    \EndFor
    \State $p \gets p + \text{``\textbf{New Intent:} "} + i$
    \State $p \gets p + \text{``\textbf{Type, Keywords}"}$
    \State \Return $p$
\EndProcedure

\Procedure{QueryLLM}{$p, LLM$}
    \State \Return \Call{LLM.Query}{$p$} 
\EndProcedure

\Procedure{ParseResponse}{$response$}
    \State $type \gets $ \Call{ExtractType}{$response$}
    \State $keywords \gets $ \Call{ExtractKeywords}{$response$}  
    \State \Return $(type, keywords)$
\EndProcedure
\end{algorithmic}
\end{algorithm}

This intent processing functionality is presented as a User Interface (UI) and intent translation application in Fig. \ref{fig2}. The job of this application is to work as a UI for a human operator and process the intents with crucial information to feed the intent validation and the meta-controller application. We further elaborate our intent processing methodology via Algorithm 1. The approach can be categorized as prompt engineering, where the prompt serves as a template for the LLM to infer the correct response based on minimal data examples.

\subsection{Predictive Intent Validation Using Transformer}

To perform intent validation, we utilize one of our previous works \cite{8} where traffic data are collected so that it can be fed to the transformer-based time series predictor for future traffic volume prediction. The traffic predictor resides in the strategic controller (named as intent validation app, as in Fig \ref{fig2}.). The forecast for traffic volume is performed for successive time intervals, with two distinct thresholds derived from historical data. The higher threshold, denoted as $Th_p$, is decided based on previous instances of heavy traffic. If the forecast traffic surpasses $Th_p$, applications like traffic steering and power allocation can be activated to enhance or sustain high throughput, depending on the intent of a human operator. The lower threshold, $Th_t$, is established based on low traffic volume. In the event that the anticipated traffic decreases to less than this threshold, an intent to increase energy efficiency can be validated. It will then lead to the initiation of energy-saving applications like cell sleeping due to the expected lower traffic levels. The algorithmic process of predictive traffic validation is presented in Algorithm 2.

\begin{algorithm}
\caption{Intent Validation Based on Traffic Prediction}
\begin{algorithmic}[1]
\State Predict traffic $T_p$ for $t+1$ and extract $i$ using Algorithm 1
\If{($T_p > Th_p$ $ || $ $T_p < Th_t$)}
    \State Check QoS profiles for the relevant traffic class $k$
    \If{$i$ impacts $Q_s$}
        \For{each QoS parameter $Q_{s_j}$ in $k_\eta$'s profile}
            \State Calculate QoS drift $I_{Q}(Q_{s_j})$ using eq. \ref{eq5}
            \If{$I_{Q}(Q_{s_j}) \neq 0$} 
                \State Intent invalid
                \State \textbf{break}
            \EndIf
        \EndFor
        \If{$I_{Q}$ for elements in $Q_s$ is 0 $||$ $A_{QoS}>D_{QoS}$}
            \State Validate intent
        \EndIf
    \EndIf
\Else
    \State Re-calculate $Th_p$ and $Th_t$
\EndIf
\State \textbf{return} Validation results
\end{algorithmic}
\end{algorithm}

\subsection{Network Optimization Using Attention-based HRL}

After intent validation, we optimize the network performance based on the operator intent using an HRL algorithm. We opt for a hierarchical Deep-Q-Network (h-DQN) having a two-level hierarchy with a meta-controller on top \cite{15} and a controller on the bottom. These two controllers are solely related to the h-DQN algorithm that we are proposing to use to manage network-optimizing applications. Meta-controller and the controller are hosted in the strategic and tactical controller, respectively. We employ an attention mechanism \cite{7} to reduce the action space for our h-DQN agent, enhancing efficiency and decision-making while expediting convergence. We can define the attention function as follows:
\begin{equation}
    \alpha(s_t, i, a) = f_\theta(s_t, i, a),
\end{equation}
where $f_\theta$ computes the attention score trained using supervised learning, to predict which actions are most likely to fulfill the operator’s intent. Once the attention scores $\alpha(s_t, i, a)$ are computed, the action space is reduced to a subset, $A_s^{att}$ containing only the most relevant actions. 
\begin{equation}
    A_s^{att} = \{ a \in A \mid \alpha(s_t, i, a) > \epsilon \},
\end{equation}
where $\epsilon$ is a threshold that filters out actions with low relevance. This subset $A_s^{att}$ is much smaller than the full action space $A$, allowing the agent to make decisions more efficiently. For each Markovian option $\omega$ \cite{7}, the attention mechanism selects only the relevant options based on the operator’s intent. Each option is defined by a policy $\pi_\omega$, a termination condition $\beta_\omega$, and an initiation set $I_s$. After reducing the action space, the agent applies its policy $\pi$ to select the best action from $A_s^{att}$:
\begin{equation}
    \pi(s_t) = \arg \max_{a \in A_s^{att}} Q(s_t, a), 
    \end{equation}
where $Q(s_t, a)$ is the action-value function that represents the expected return from taking action $a$ in state $s_t$.

Meta-controller in h-DQN can take in a network state (e.g., traffic class) and a goal (desired change of a performance metric extracted from an intent) to achieve. The controller in the lower level takes the action of choosing an application or combinations of them based on the state and the goal. The meta-controller is designed as a top-level application in the strategic controller (see Fig. \ref{fig2}). Applications are controlled by the meta-controller that can directly optimize system performance. At this stage of the work, we define the following MDP for the h-DQN to solve the problem formulated in (\ref{eq4}). 

\begin{itemize}
    \item \textbf{State and Action:} The set of states consists of traffic flow types of different users in the network. $S=\{T_1,..,T_2,...,T_3,...,T_4,..\}$.  Both meta-controller and controller share the same states. Selecting network optimizing applications, or combinations of them is considered as actions to be performed by the controller which is defined as  \{$A=A_{App1}, A_{App1,2},...., A_{AppN}\}$.
    \item \textbf{Intrinsic reward:} The intrinsic reward function ($r_{in}$) for the controller is: $r_{in}=C_\rho-\varpi \gamma_{q_s}$ (see (\ref{eq4}) for more details.).
    \item \textbf{Goal for the controller:} Increased level of a performance metric that can satisfy operator intent is passed to the controller as goals. For example, $G=\{TP_1, TP_2,..., TP_n\}$ for throughput increasing intents.
    \item \textbf{Extrinsic reward:} Summation of the intrinsic reward over $\tau$ steps.
\end{itemize}

We have used five applications in total that are controlled by the meta-controller. They are traffic steering \cite{10}, cell sleeping \cite{1}, beamforming \cite{1}, power allocation \cite{19}, and energy efficient handover management \cite {16}. All of these applications are based on DRL. The traffic steering application is engineered to concurrently uphold QoS across various traffic types by employing a steering mechanism to ensure optimal performance in terms of delay and throughput. The cell sleeping app (works on a longer time scale in the strategic controller) aims to reduce network power consumption by deactivating idle or less busy BSs based on traffic load and queue length. The beamforming application uses UE coordinates to sort out the steering angle, and the array steering vector corresponding to the respective element in the codebook. The power allocation application tries to maximize the total throughput and lastly, the handover management app is specifically tailored with DRL to achieve energy efficiency via optimally tailored handover policies. The process of selecting and orchestrating apps for network optimization using HRL can be summarized as follows: 

\begin{itemize}
    \item \textbf{Step 1:} An input is provided by the human network operator associated with a performance metric.
    \item \textbf{Step 2:} Intent is processed and crucial information like intent type: “energy-efficiency/ throughput/ Delay”, and magnitude of change (in percentage) are extracted using pre-trained LLM.
    \item \textbf{Step 3:} Extracted information is passed to the intent validation application.
    \item \textbf{Step 4:} With a valid intent, a target associated with the performance metrics (goals) and feasible action set for the intent are provided to the controller by the meta-controller.
    \item \textbf{Step 5:} The controller selects an application or a combination of them to reach the target performance as close as possible.
    \item \textbf{Step 6:}  Selected applications optimize the performance of the network as a response to the intent of the operator. 
\end{itemize}

\section{Performance Evaluation}
\label{s5}
In this section, first, we introduce the simulation settings followed by the simulation results showing the effectiveness and superiority of the proposed method.

\subsection{Simulation setup}

\begin{figure}[!t]
\centerline{\includegraphics[width=0.8\linewidth]{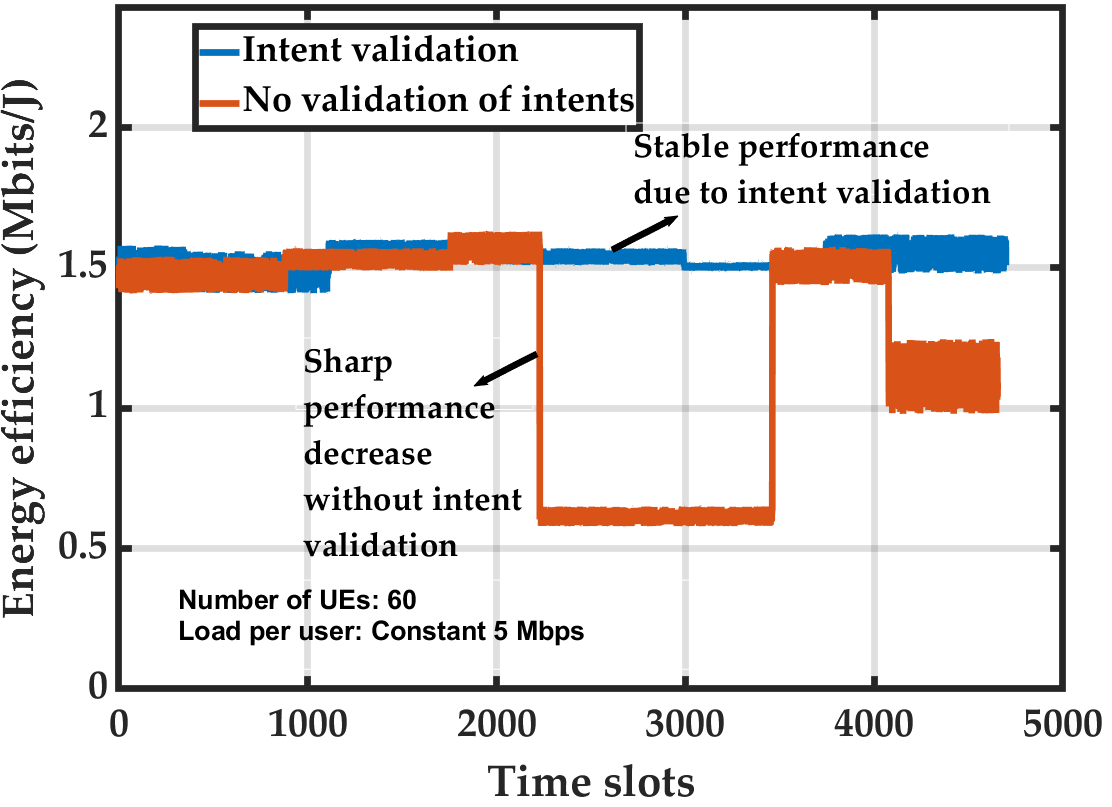}}
\caption{Impact of intent validation on energy efficiency.}
\label{fig4}
\vspace{-1.2em}
\end{figure}

\begin{figure}[!t]
\centerline{\includegraphics[width=0.8\linewidth]{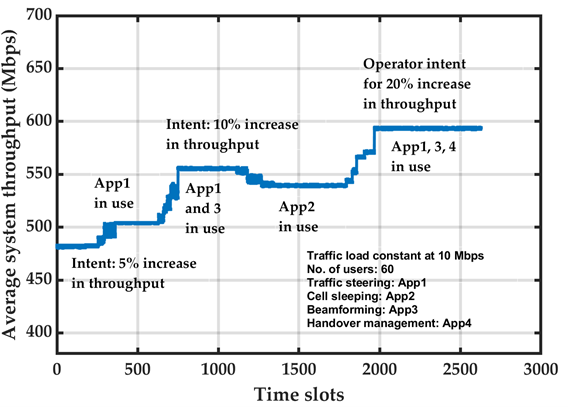}}
\caption{Impacts of operator intents on throughput.}
\label{fig5}
\vspace{-1.2em}
\end{figure}

The simulation setup for this study features a macro cell surrounded by densely deployed small cells in a multi-RAT (multi-radio access technology) environment, accommodating a total of 60 users. For the 5G NR configuration, a bandwidth of 60 MHz is used, with carrier frequencies of 3.5 GHz and 30 GHz, a subcarrier spacing of 15 KHz, and a maximum transmission power of 43 dBm, based on the 3GPP urban micro channel model \cite{18}. The LTE setup, on the other hand, employs a 40 MHz bandwidth, an 800 MHz carrier frequency, a 15 KHz subcarrier spacing, and a maximum transmission power of 38 dBm, following the 3GPP urban macro channel model \cite{18}.

There are four traffic types: video, gaming, voice, and Ultra Reliable Low Latency Communications (URLLC), each characterized by specific inter-arrival times for packets: 12.5 ms, 40 ms, 20 ms, and 0.5 ms, respectively. These packet inter-arrival times follow different distributions—Pareto for video, Uniform for gaming, and Poisson for both voice and URLLC. In the following section, we present results using shorthand identifiers for applications: App1 (traffic steering), App2 (cell sleeping), App3 (beamforming), App4 (power allocation), and App5 (handover management).

Table \ref{tab1} provides a summary of the tunable parameters for both the Autoformer (used for traffic prediction) and the RL methods (including the HRL control algorithm and DRL-based network applications).

\begin{table}[htp]
    \centering
    \caption{Parameter settings for Autoformer and RL}
    \begin{tabular}{|l|l|}
         \hline
         \textbf{\underline{Autoformer}} & \\
         Learning rate $(\alpha)$, Batch size & $10^{-4}$, $32$ \cite{8} \\
         Model architecture & 2 encoder layers, 1 decoder layer\\
         \hline
         \textbf{\underline{HRL parameters (h-DQN)}} & \\
         Exploration Probability ($\epsilon_2$) & $1$, annealed to $0.1$ \cite{15} \\
         Learning rate ($\alpha$), discount factor $\gamma$ & $0.00025$, $0.99$  \cite{15} \\
         \hline
         \textbf{\underline{RL paramters (applications)}} & \\
         Batch size, Initial exploring steps & $32$, $3000$ \\
         Learning rate ($\alpha$), discount factor $\gamma$ & $0.5$, $0.9$ \\
         \hline
        
    \end{tabular}
    \label{tab1}
\end{table}
\vspace{-5pt}

\subsection{Simulation Results}

The first simulation result is associated with intent validation. When there is no intent validation, conflicting intents from an operator can severely degrade performance. Fig. \ref{fig4} presents such an example where we have a traffic load of 5Mbps. At $2200^{th}$ time slot, there is a sharp decrease in energy efficiency. This is because of the unwanted intent of increasing throughput even though the cell sleeping application is active due to low traffic demand. An intent of increasing throughput would result in activating more cells unnecessarily and decrease overall energy efficiency. Similar impacts can be observed for other KPIs like delay, and throughput also.

As shown in Fig. \ref{fig5}, the human operator's goal to “increase throughput” leads to multiple application selection. For instance, aiming for a $5\%$ throughput increase results in a noticeable surge after several time slots due to App1. Aiming for a $10\%$ increase involves both App1 and App3. Conversely, reducing power consumption significantly decreases throughput around the $1250^{th}$ time slot, reflecting the deactivation of App1 and App3 and the activation of App2. When the operator aims for a $20\%$ increase, App1, App3, and App4 together produce a sharp throughput rise. 

\begin{figure}[!t]
\centerline{\includegraphics[width=1\linewidth]{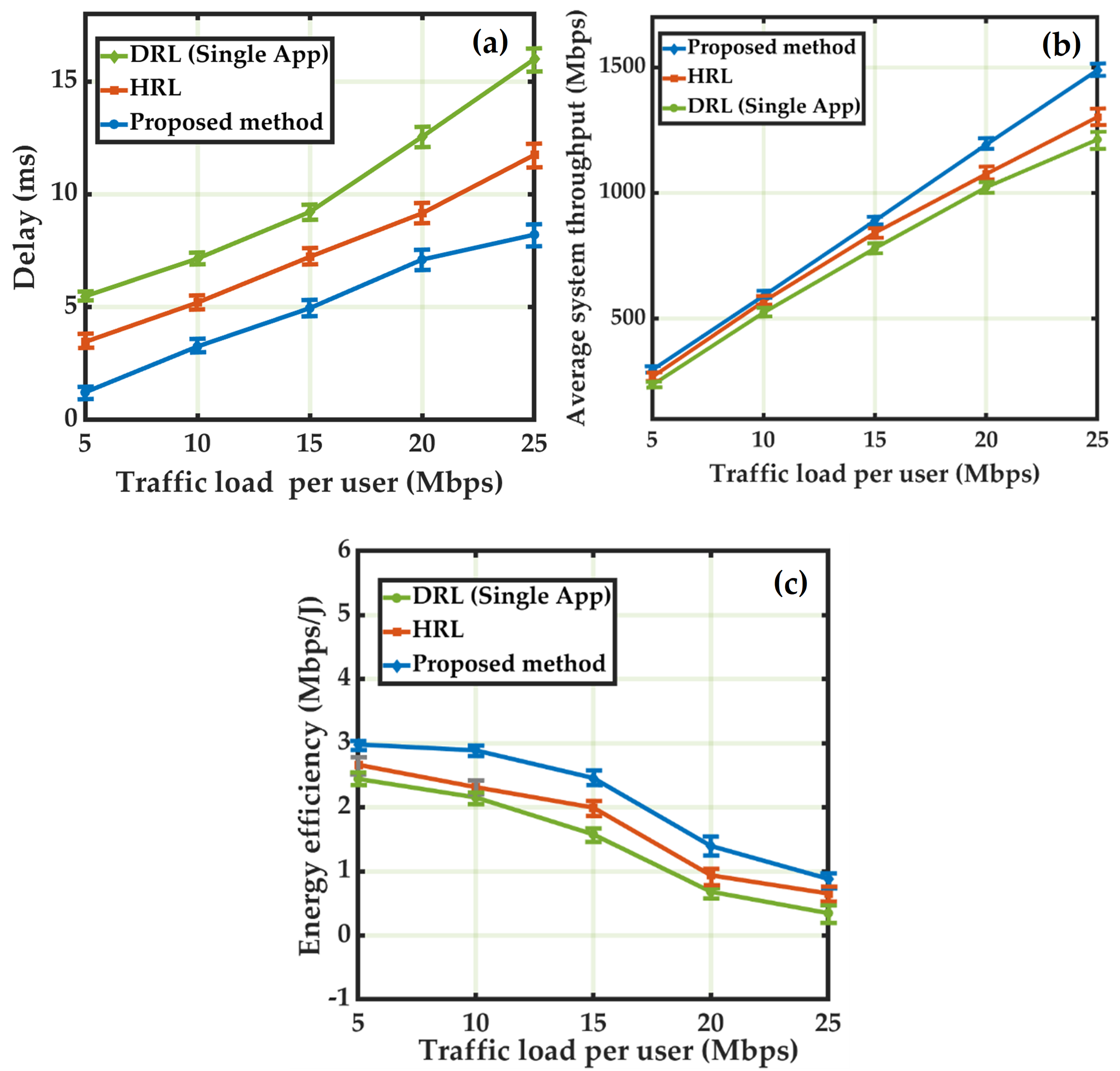}}
\caption{Performance analysis of the proposed method: (a) network delay, (b) throughput, and (c) energy efficiency.}
\label{fig3}
\vspace{-1.2em}
\end{figure}

The proposed method's effectiveness is compared with two baselines: an HRL-based application initiation without an attention mechanism or intent validation, and a single DRL-based application scenario. The proposed method employs a hard attention mechanism, focusing on the subset of applications most relevant for the current traffic type and operator's intent. This focus reduces conflicting actions and shows superior performance over the HRL baseline (improvements of $12.02\%$ in throughput, $26.5\%$ in delay, and $17.1\%$ in energy efficiency) and the DRL baseline (improvements of $17.25\%$, $48.6\%$, and $39.3\%$ in the same metrics), as observed in Fig. \ref{fig3}.

\section{Conclusions}
\vspace{-5pt}
\label{s6}  
In this paper, first, we process intents using an LLM and extract crucial information regarding which performance metric to optimize and by how much. Next, a Transformer-based intent validation technique has been used to rule out intents that conflict with the contemporary network state to avoid performance degradation. Lastly, we use an attention-based HRL to initiate and orchestrate applications for performance optimization. The proposed method outperforms the HRL baseline in throughput, delay, and energy efficiency by $12.02\%$, $26.5\%$, and $17.1\%$ respectively. Additionally, compared to the DRL baseline, it achieves superior results across the same performance metrics, with enhancements of $17.25\%$, $48.6\%$, and $39.3\%$ respectively. In the future, we plan to perform highly complex intent translations from the human network operator.

\section*{Acknowledgement}
\vspace{-5pt}
This work has been supported by MITACS, Ericsson Canada, and the NSERC Canada Research Chairs program.

\bibliographystyle{IEEEtran}
\bibliography{reference.bib}
\end{document}